\documentclass[12pt]{article}

\usepackage{latexsym,amsmath,amssymb,amsthm,amsfonts,graphicx,natbib,breqn,url}
\usepackage{epsfig}
\usepackage{bm,color}
\usepackage{authblk}
\usepackage{setspace}
\usepackage{subcaption}

\newcommand*\patchAmsMathEnvironmentForLineno[1]{%
      \expandafter\let\csname old#1\expandafter\endcsname\csname #1\endcsname
      \expandafter\let\csname oldend#1\expandafter\endcsname\csname end#1\endcsname
      \renewenvironment{#1}%
         {\linenomath\csname old#1\endcsname}%
         {\csname oldend#1\endcsname\endlinenomath}}%
    \newcommand*\patchBothAmsMathEnvironmentsForLineno[1]{%
      \patchAmsMathEnvironmentForLineno{#1}%
      \patchAmsMathEnvironmentForLineno{#1*}}%
    \AtBeginDocument{%
    \patchBothAmsMathEnvironmentsForLineno{equation}%
    \patchBothAmsMathEnvironmentsForLineno{align}%
    \patchBothAmsMathEnvironmentsForLineno{flalign}%
    \patchBothAmsMathEnvironmentsForLineno{alignat}%
    \patchBothAmsMathEnvironmentsForLineno{gather}%
    \patchBothAmsMathEnvironmentsForLineno{multline}%
    }
\usepackage[mathlines,displaymath]{lineno}
\usepackage{float} 
\floatstyle{plain}
\newfloat{Algorithm}{thp}{lop}
\floatname{Algorithm}{Algorithm}
\newenvironment{wideenumerate}{\enumerate\addtolength{\itemsep}{5pt}}{\endenumerate}

\include{def}
\def\dispmuskip{\thinmuskip= 3mu plus 0mu minus 2mu \medmuskip=  4mu plus 2mu minus 2mu \thickmuskip=5mu plus 5mu minus 2mu}
\def\textmuskip{\thinmuskip= 0mu                    \medmuskip=  1mu plus 1mu minus 1mu \thickmuskip=2mu plus 3mu minus 1mu}
\def\beq{\dispmuskip\begin{equation}}    \def\eeq{\end{equation}\textmuskip}
\def\beqn{\dispmuskip\begin{displaymath}}\def\eeqn{\end{displaymath}\textmuskip}
\def\bea{\dispmuskip\begin{eqnarray}}    \def\eea{\end{eqnarray}\textmuskip}
\def\bean{\dispmuskip\begin{eqnarray*}}  \def\eean{\end{eqnarray*}\textmuskip}

\begin{document}

\title{Gaussian variational approximation with a factor covariance structure}
\date{\empty}

\author[1]{Victor M.-H. Ong}
\author[1]{David J. Nott\thanks{Corresponding author:  standj@nus.edu.sg}}
\author[2]{Michael S. Smith}
\affil[1]{Department of Statistics and Applied Probability, National University of Singapore, Singapore 117546}
\affil[2]{Melbourne Business School, University of Melbourne, 200 Leicester Street, Carlton, VIC, 3053.}

\maketitle

\vspace{-0.3in}

\begin{abstract}
Variational approximation methods have proven to be useful for scaling Bayesian computations to large data sets and highly parametrized models.  
Applying variational methods involves solving an optimization problem, and recent research in this area has focused
on stochastic gradient ascent methods as a general approach to implementation.
Here variational approximation is considered for a posterior distribution in high dimensions using a Gaussian approximating family.
Gaussian variational approximation with an unrestricted covariance matrix can be computationally burdensome in many problems because the number
of elements in the covariance matrix increases quadratically with the dimension of the model parameter.  
To circumvent this problem, low-dimensional factor covariance structures are considered.  
General stochastic gradient approaches to efficiently perform the optimization are described, with gradient estimates
obtained using the so-called ``reparametrization trick".  The end result is a flexible and efficient approach to high-dimensional
Gaussian variational approximation, which we illustrate using eight real datasets.


\end{abstract}

\section{Introduction}\label{sec:Intro}

Variational approximation methods are a promising approach to scalable approximate Bayesian inference in the case of large
data sets and highly parametrized models.  However, if the variational approximation takes the form of a multivariate Gaussian distribution with an unrestricted covariance matrix, 
it is difficult to perform variational inference with a high-dimensional
parameter because the number of elements in the covariance matrix increases quadratically with the parameter dimension.  
Hence, in the context of Gaussian variational approximation, it is important to find parsimonious but flexible ways of parametrizing the covariance matrix.  
The contribution of the present paper is to develop general 
methods for Gaussian variational approximation when the covariance matrix has a factor structure.  
By general here, we mean that the methods do not require any special structure for the prior and likelihood function.  
A key feature of our approach is that we obtain efficient
gradient estimates for a stochastic gradient ascent optimization procedure using the so-called ``reparametrization trick".  This leads to a flexible and computationally attractive approach to high-dimensional
Gaussian variational approximation. 

Let $\theta$ be a continuous parameter of dimension $m$, and consider Bayesian inference with a prior density $p(\theta)$ and likelihood $p(y|\theta)$. 
Write the posterior density as $p(\theta|y)$, and to simplify notation later write $h(\theta)=p(\theta)p(y|\theta)$, so that $p(\theta|y)\propto h(\theta)$.  
Variational approximation methods \citep{Attias1999,Jordan1999,Winn2005,Ormerod2010} 
provide approximate methods for performing Bayesian calculations having reduced
computational demands compared to exact methods such as Markov chain
Monte Carlo (MCMC).  In a variational approach it is assumed that the posterior density can be approximated by a member of some tractable family of approximations, 
with typical element $q_\lambda(\theta)$ say, where $\lambda$ are variational parameters to be chosen indexing different members of the family.  
Writing $p(y)=\int p(\theta)p(y|\theta)d\theta$ for the marginal likelihood of $y$, the following identity holds, for any $q_\lambda(\theta)$:
\begin{align}
  \log p(y) = & \int \log\frac{p(\theta)p(y|\theta)}{q_\lambda(\theta)}q_\lambda(\theta)d\theta+\text{KL}(q_\lambda(\theta)||p(\theta|y)),  \label{identity}
\end{align}
where 
\begin{align*}
  \mbox{KL}(q_\lambda(\theta)|| p(\theta|y)) = & \int \log \frac{q_\lambda(\theta)}{p(\theta|y)} q_\lambda(\theta) d\theta
\end{align*}
is the Kullback-Leibler divergence from $q_\lambda(\theta)$ to $p(\theta|y)$.  
Derivation of equation (\ref{identity}) can be found, for example, in \citet[p. 42]{Ormerod2010}.  
We denote the expectation with respect to $q_\lambda(\theta)$ as $E_q(\cdot)$.
Because the Kullback-Leibler divergence is non-negative, from equation (\ref{identity}),
\begin{align}
 {\cal L}(\lambda)= & \int \log\frac{p(\theta)p(y|\theta)}{q_\lambda(\theta)}q_\lambda(\theta)d\theta= E_q(\log h(\theta)-\log q_\lambda(\theta))  \label{lowerbound1}
\end{align}
is a lower bound on $\log p(y)$, called the variational lower bound.  
The Kullback-Leibler divergence is one useful measure of the quality of the approximation of the true posterior by $q_\lambda(\theta)$, and we choose
$\lambda$ so that the approximation is optimal.  
The lower bound will be tight when $q_\lambda(\theta)$ is equal
to the true posterior, since the Kulback-Leibler divergence is zero in this case.  Because the left hand side of (\ref{identity}) doesn't depend on the variational
parameters, minimizing the Kullback-Leibler divergence $\mbox{KL}(q_\lambda(\theta)||p(\theta|y))$ with respect to $\lambda$ 
is equivalent to maximizing ${\cal L}(\lambda)$ with respect to $\lambda$.    Therefore, maximizing ${\cal L}(\lambda)$ with respect to $\lambda$ 
provides the best approximation to our posterior distribution within the approximating class in the Kullback-Leibler sense.
In this manuscript we will be concerned with the situation where $q_\lambda(\theta)$ is multivariate normal, so without
any further restriction the variational parameters $\lambda$ consist of both the mean vector and distinct elements of the covariance matrix of the normal variational posterior.
As mentioned above, a full normal variational approximation is difficult to work with in high dimensions.  
Assuming a diagonal covariance structure is one possible simplification,
but this loses any ability to represent dependence in the posterior distribution.  

Various suggestions in the literature exist for parsimonious ways to parametrize covariance matrices in Gaussian variational approximations, while retaining some 
representation of dependence between the model parameters.
\citet{Opper2009} note that with a Gaussian prior and a factorizing likelihood,
the optimal Gaussian variational distribution can be specified in terms of a much reduced set of variational parameters. 
\citet{Challis2013} consider posterior distributions which can be expressed as 
a product of a Gaussian factor and positive potential, and consider banded Cholesky, Chevron Cholesky and subspace Cholesky approximations.  They are also able
to prove concavity of the variational lower bound in this setup.  \cite{Titsias2014} consider both full and diagonal covariance structures with the covariance matrix
parametrized in terms of the Cholesky factor, where stochastic gradient variational Bayes methods are used to do the optimization in quite a general way.  
Efficient gradient estimates are constructed using the so-called ``reparametrization trick" \citep{Kingma2013,rezende+mw14}.  
\citet{Kucukelbir2016} consider both unrestricted and diagonal covariance matrices, as well as marginal transformations to improve normality, working in an automatic differentiation
environment and using similar gradient estimates to \cite{Titsias2014}.  
\citet{Salimans2013} consider a variety of stochastic gradient optimization approaches for learning exponential family type approximations or hierarchical extensions
of such approximations.  In the Gaussian case, they mostly consider parametrizations of the covariance matrix in terms of the precision matrix, and
are able to exploit sparsity of Hessian matrices for the joint model in their computations, with such sparsity being related to conditional independence structure.  
As well as their algorithm using the Hessian, they also provide algorithms that require only computation of first order derivatives.
\citet{Archer2016} consider Gaussian variational approximation in the context of smoothing for state space models.  They parametrize the variational optimization in terms
of a sparse precision matrix, and exploit the way that this leads to a sparse Cholesky factor in random variate generation from their variational posterior
distribution.  The blocks of the mean vector and non-zero blocks of the precision matrix are parametrized in terms of global parameters that relate 
them to local data -- an example of so-called amortized variational inference -- which was also introduced in \citet{Kingma2013}.  
\citet{Tan2016} parametrize the variational optimization directly in terms of the Cholesky factor of the precision matrix and impose sparsity on the
Cholesky factor that reflects conditional independence relationships.  They show how the sparsity can be exploited in the computation of gradients
with the reparametrization trick.  

In the above work the approximations considered either require some special structure of the model (such as conditional independence structure, Gaussian priors or a factorizing likelihood), 
do not scale well to high dimensions, or are inflexible in the kinds of dependence they can represent accurately.  
The goal of the present work is to consider a general method for Gaussian variational approximation, where the covariance matrix is parametrized
in terms of a factor structure.  Factor models are well known to be a very successful approach to modelling high-dimensional covariance matrices in many circumstances
\citep{Bartholomew2011}.  By assuming a factor stucture the number of variational parameters is reduced considerably when the number of factors is much
less than the full dimension of the parameter space.  
Such a parsimonious approximation has strong potential in certain applications.  For example, in random effects models dependence among the high-dimensional vector of random effects
can often be explained by their shared dependence on just a small number of global parameters.   We demonstrate this later for a mixed effects logistic regression model.
Gaussian variational approximations with a factor covariance structure have been considered previously by
\citet{Barber1998} and \citet{Seeger2000}.  However, these authors consider models with special structure in which the variational lower bound can be evaluated
analtyically, or using one-dimensional numerical quadrature.  In contrast, here we consider approaches to performing the required variational
optimization without requiring any special structure for the prior or a factorizing likelihood.  In independent work \citet{Miller2016} have recently also suggested
the use of factor parametrizations of covariance structure in Gaussian variational approximation, using stochastic gradient methods and the reparametrization trick for gradient estimation.  
However, their focus is on building mixture of Gaussian variational approximations using a boosting perspective and they do not give expressions for the gradient estimates for the 
Gaussian factor components or the derivation of such results.

In the next section we briefly introduce the main ideas of stochastic gradient variational Bayes.  
Section 3 then gives details of our stochastic gradient ascent algorithm for 
optimization of the variational parameters in a Gaussian approximation with factor covariance structure.  
Efficient gradient estimation based on the reparametrization trick is developed, and we show that matrix computations in the gradient calculations can be done efficiently using
the Woodbury formula.  Derivation of the gradient experssions are given in the Appendix.  Section 4 illustrates the advantages of the method by applying it to eight examples and Section 5 concludes.

\section{Stochastic gradient variational Bayes}

We note that ${\cal L}(\lambda)$ in (\ref{lowerbound}) is defined in terms of an expectation, and when this cannot be evaluated in closed form a number of authors
\citep{ji+sw10,Paisley2012,nott+tvk12,Salimans2013,Kingma2013,rezende+mw14,Hoffman2013,Ranganath2014,Titsias2015}
have suggested optimizing ${\cal L}(\lambda)$ using stochastic gradient ascent methods \citep{Robbins1951}.  
If ${\cal L}(\lambda)$ is the objective function to be optimized, $\nabla_\lambda {\cal L}(\lambda)$ is its gradient, and 
$\widehat{\nabla_\lambda {\cal L}(\lambda)}$ is an unbiased estimate of the gradient, then the basic form of a stochastic gradient ascent optimization is as follows.
After choosing an initial value $\lambda^{(0)}$ for the variational parameters $\lambda$,  for $t=0,1,\dots$ perform the update
\begin{align*}
 \lambda^{(t+1)} & =\lambda^{(t)}+\rho_t \widehat{\nabla_\lambda {\cal L}(\lambda^{(t)})}
\end{align*}
until a stopping condition is satisfied. Here, $\rho_t$, $t\geq 0$, is a sequence of learning rates, typically chosen to satisfy the Robbins-Monro conditions
\citep{Robbins1951} $\sum_t \rho_t=\infty$ and $\sum_t \rho_t^2<\infty$.  Convergence of the sequence $\lambda^{(t)}$ will be to a local
optimum under regularity conditions \citep{bottou10}.  In practice it is important to consider adaptive learning rates, and in our later examples we implement
the ADADELTA approach \citep{Zeiler2012}, although there is a large literature on different adaptive choices of the learning rates.

The references given above differ in the way
that the unbiased gradient estimates $\widehat{\nabla_\lambda {\cal L}(\lambda)}$ are constructed, and the variance reduction methods employed.  
Reducing the variance of the gradient estimates is important because this affects the stability and speed of convergence of the algorithm.
Differentiating directly under the integral sign in (\ref{lowerbound1}) and using
the fact that $E_q(\nabla_\lambda \log q_\lambda(\theta))=0$ (the so-called log-derivative trick) and some simple algebra, the gradient is
\begin{align}
 \nabla_\lambda {\cal L}(\lambda) = & E_q(\nabla_\lambda \log q_\lambda(\theta)(\log h(\theta)-\log q_\lambda(\theta))). \label{logdtrick}
\end{align}
Since this is an expectation with respect to $q_\lambda(\theta)$, it is easy to estimate (\ref{logdtrick}) unbiasedly using samples
from $q_\lambda(\theta)$, provided that sampling from $q_\lambda(\theta)$ is possible.  In 
large data sets this can also be combined with unbiased estimation of $\log h(\theta)$ using subsampling of terms in the log-likelihood 
(so-called doubly stochastic variational inference, see \citet{Salimans2013,Kingma2013} and \citet{Titsias2014} for example).

In practice, even with sophisticated variance reductions it is often found that derivatives obtained from (\ref{logdtrick}) can have high variance, 
and an alternative approach was considered by \citet{Kingma2013} and \citet{rezende+mw14}, which they have called the reparametrization trick.  
To apply this approach, we need to be able
to represent samples from $q_\lambda(\theta)$ as $\theta=t(\epsilon,\lambda)$, where $\epsilon$ is a random vector with a fixed density $f(\epsilon)$ that does not
depend on the variational parameters. In particular, in the case of a Gaussian variational distribution parametrized in terms of a mean vector $\mu$ and the Cholesky
factor $C$ of its covariance matrix, we can write $\theta=\mu+C\epsilon$, where $\epsilon\sim N(0,I)$.  Then 
\begin{align}
  {\cal L}(\lambda)= & E_q(\log h(\theta)-\log q_\lambda(\theta)) \nonumber \\
  = & E_f(\log h(t(\epsilon,\lambda))-\log q_\lambda(t(\epsilon,\lambda))), \label{repartrick}
\end{align}
where we have written $E_f(\cdot)$ to denote expectation with respect to $f(\cdot)$.  
Differentiating under the integral sign in (\ref{repartrick}) gives an expectation with respect to $f(\cdot)$ that can be estimated unbiasedly based on samples from $f(\cdot)$.  
Because of the reparametrization in terms of $\epsilon$, the variational parameters have been moved inside the function $h(\cdot)$ so that when we differentiate
(\ref{repartrick}) we are using derivative information from the target posterior density.  In practice it is found that when the reparametrization trick can be applied,
it helps greatly to reduce the variance of gradient estimates.  

\section{Approximation with factor covariance structure}\label{sec:factor}

In our factor parametrization of the variational distribution it is assumed that $q_\lambda(\theta)=N(\mu,BB^T+D^2)$ where $\mu$ is the mean vector, 
$B$ is a $m\times p$ full rank matrix with $p<<m$ and $D$ is a diagonal matrix with diagonal elements $d=(d_1,\dots,d_m)$.  Without further restrictions $B$ is unidentified, and 
here we impose the restriction that the upper triangle of $B$ is zero, similar to \citet{Geweke1996}.  For uniqueness we may also wish to impose 
the restriction on the leading diagonal elements $B_{ii}>0$, 
but we choose not to do this in the present work as it does not pose any problem for the variational optimization and it is more convenient to work with the 
unconstrained parametrization.  Note that we can draw $\theta\sim N(\mu,BB^T+D^2)$ by first drawing $(z,\epsilon)\sim N(0,I)$ (where $z$ is $p$-dimensional and
$\epsilon$ is $m$ dimensional) and then calculating
$\theta=\mu+Bz+d\circ \epsilon$, where $\circ$ denotes the Hadamard (element by element) product of two random vectors.
This will be the basis for our application of the reparametrization trick, and 
also makes explicit the intuitive idea behind factor models, which is that correlation among the components may be explained in terms of a smaller
number of latent variables ($z$ in this case) which influence all the components, with component specific ``idiosyncratic" variance being captured through 
the additional independent error term $d\circ\epsilon$.  

We now explain how to apply the reparametrization trick of \citet{Kingma2013} and \citet{rezende+mw14} 
to obtain efficient gradient estimates for stochastic gradient variational inference in this setting.  
Write $f(z,\epsilon)$ for the $N(0,I)$ density of $(z,\epsilon)$ in the generative representation of $q_\lambda(\theta)$ described above.
The lower bound is an expectation with respect to $q_\lambda(\theta)$, but applying the reparametrization trick gives
\begin{align}
 {\cal L}(\lambda) = & E_f(\log h(\mu+Bz+d\circ \epsilon)+\frac{m}{2}\log 2\pi+\frac{1}{2}\log |BB^T+D^2|  \nonumber \\
 & \;\;\;\;+\frac{1}{2}(Bz+d\circ\epsilon)^T (BB^T+D^2)^{-1}(Bz+d\circ\epsilon)). \label{lowerbound}
\end{align}
We give some expressions for the components of $\nabla_\lambda {\cal L}(\lambda)$ obtained from differentiating in (\ref{lowerbound}) under
the integral sign, but first we need some notation. 
For a matrix $A$, we write $\mbox{vec}(A)$ for the vector obtained by stacking the columns of $A$ one underneath the other
as we go from left to right.  We will not require that $A$ be a square matrix.  
We write $\mbox{vec}^{-1}(\cdot)$ for the inverse operation (where in what follows the dimensions of the resulting matrix will be clear from the context and we will not
make this explicit in the notation).
Also, for a vector $x$ and real valued function $g(x)$, we write $\nabla_x g(x)$ for the gradient vector, written as a column vector, and 
for a matrix $A$ and real-valued function $g(A)$ we define $\nabla_A g(A)=\mbox{vec}^{-1}(\nabla_{\mbox{vec}(A)} g(A))$ so that $\nabla_A g(A)$ is a matrix
of the same dimensions as $A$.  Also, we write $\mbox{diag}(Z)$ for the vector of diagonal entries of the square matrix $Z$.

With this notation, it is shown in the Appendix that,
\begin{align}
  \nabla_\mu {\cal L}(\lambda)= & E_f(\nabla_\theta \log h(\mu+Bz+d\circ\epsilon)), \label{gradmu}
\end{align}
\begin{align}
  \nabla_B {\cal L}(\lambda) = & E_f(\nabla_\theta \log h(\mu+Bz+d\circ\epsilon)z^T+(BB^T+D^2)^{-1}B+(BB^T+D^2)^{-1}(Bz+d\circ\epsilon)z^T \nonumber \\
 & -(BB^T+D^2)^{-1}(Bz+d\circ\epsilon)(Bz+d\circ\epsilon)^T(BB^T+D^2)^{-1}B) \label{gradB1}
\end{align}
and
\begin{align}
  \nabla_d {\cal L}(\lambda) = & E_f(\mbox{diag}(\nabla_\theta \log h(\mu+Bz+d\circ\epsilon)\epsilon^T+(BB^T+D^2)^{-1}D+(BB^T+D^2)^{-1}(Bz+d\circ\epsilon)\epsilon^T \nonumber \\
 & -(BB^T+D^2)^{-1}(Bz+d\circ\epsilon)(Bz+d\circ\epsilon)^T(BB^T+D^2)^{-1}D)). \label{gradd1}
\end{align}
However, also noting that the second and fourth terms in (\ref{gradB1}) and (\ref{gradd1}) are equal after taking expectations,
\begin{align}
  \nabla_B {\cal L}(\lambda) = & E_f(\nabla_\theta \log h(\mu+Bz+d\circ\epsilon)z^T+(BB^T+D^2)^{-1}(Bz+d\circ\epsilon)z^T)  \label{gradB}
\end{align}
and
\begin{align}
  \nabla_d {\cal L}(\lambda) = & E_f(\mbox{diag}(\nabla_\theta \log h(\mu+Bz+d\circ\epsilon)\epsilon^T+(BB^T+D^2)^{-1}(Bz+d\circ\epsilon)\epsilon^T).  \label{gradd}
\end{align}
Estimating the expectations in these gradient expressions based on one or more samples from $f$ gives an unbiased estimate
$\widehat{\nabla_\lambda {\cal L}(\lambda)}$ of $\nabla_\lambda {\cal L}(\lambda)$.  This can be used in a stochastic gradient ascent algorithm for optimizing the
lower bound, resulting in Algorithm \ref{Alg1}.  Use of expressions (\ref{gradB}) and (\ref{gradd}) is preferable to (\ref{gradB1}) and (\ref{gradd1}).  This is 
because near the mode of ${\cal L}(\lambda)$, if the true posterior is Gaussian with the assumed covariance structure holding,
then the gradient estimates based on (\ref{gradB}) and (\ref{gradd}) for just a single sample tend to zero, whereas the alternative expressions (\ref{gradB1}) and (\ref{gradd1}) 
add noise.  Specifically, if $h(\theta)$ is proportional to $q_\lambda(\theta)$ at the modal $\lambda$ value, then by differentiating the expression for $\log q_\lambda(\theta)$
we obtain
\begin{align*}
 \nabla_\theta \log h(\mu+Bz+d\circ\epsilon)=-(BB^T+D^2)^{-1}(Bz+d\circ\epsilon)
\end{align*}
which shows that a gradient estimate based on a single sample of $f$ using (\ref{gradB}) and (\ref{gradd}) will be zero at the mode.
Similar points are discussed in \citet{Salimans2013}, \citet{Han2016} and \citet{Tan2016} in other contexts and we use the gradient estimates
based on (\ref{gradB}) and (\ref{gradd}) and a single sample from $f$ in the examples.   
\begin{Algorithm*}[htbp]
\centering
\parbox{0.8\textwidth}{
\hrule
\vspace{1mm}
Initialize $\lambda=\lambda^{(0)}=(\mu^{(0)},B^{(0)},d^{(0)})$, $t=0$.\\
Cycle
\begin{wideenumerate}
\item Generate $(\epsilon^{(t)},z^{(t)})\sim N(0,I)$
\item Construct unbiased estimates $\widehat{\nabla_\mu {\cal L}(\lambda)}$, $\widehat{\nabla_B {\cal L}(\lambda)}$, $\widehat{\nabla_d {\cal L}(\lambda)}$
of the gradients (\ref{gradmu}), (\ref{gradB}) and (\ref{gradd}) at $\lambda^{(t)}$ where the expectations are approximated from the single sample $(\epsilon^{(t)},z^{(t)})$.  
\item Set adaptive learning rate $\rho^{(t)}$ using ADADELTA or other method.
\item Set $\mu^{(t+1)}=\mu^{(t)}+\rho_t \widehat{\nabla_{\mu} {\cal L}(\lambda^{(t)})}$.
\item Set $B^{(t+1)}=B^{(t)}+\rho_t \widehat{\nabla_{B} {\cal L}(\lambda^{(t)})}$ for elements of $B^{(t+1)}$ on or below the diagonal, with the upper triangle of $B^{(t+1)}$ fixed at zero.
\item Set $d^{(t+1)}=d^{(t)}+\rho_t \widehat{\nabla_{d} {\cal L}(\lambda^{(t)})}$.
\item Set $\lambda^{(t+1)}=(\mu^{(t+1)},B^{(t+1)},d^{(t+1)})$, $t\rightarrow t+1$.  
\end{wideenumerate}
until some stopping rule is satisfied
\vspace{1mm}
\hrule}
\caption{Gaussian variational approximation algorithm with factor covariance structure.\label{Alg1}}
\end{Algorithm*} 

At first sight it may seem that computing the gradient estimates based on (\ref{gradmu}), (\ref{gradB}) and (\ref{gradd}) is difficult when $\theta$ is high-dimensional because
of the inverse of the dense $m\times m$ matrix $(BB^T+D^2)$ in these expressions.  However, note that by the Woodbury formula we have
\begin{align*}
  (BB^T+D^2)^{-1} = & D^{-2}-D^{-2}B(I+B^TD^{-2}B)^{-1}B^T D^{-2}
\end{align*}
and  that on the right hand side the matrix $(I+B^TD^{-2}B)$ is $p\times p$ with $p<<m$ and $D$ is diagonal.  So any computation involving
$(BB^T+D^2)^{-1}$ or solutions of linear systems in $(BB^T+D^2)$ can be done efficiently in terms of both memory and computation time.  

\section{Examples}\label{sec:examples}

We now demonstrate the advantages of our proposed method, which we call variational approximation with factor covariance structure (VAFC), for the case of a logistic
regression model.  
Suppose we are given a dataset with response $y_i \in \{-1,1 \}$ and covariates $\tilde{x}_i \in \mathcal{R}^q$ for $i = 1,...,n$.  For a logistic regression, the likelihood is $p(y | \theta) = \prod_{i=1}^n 1/(1 + e^{- y_i x_i^T \theta})$ where $x_i = [1 \:\: \tilde{x}_i^T]^T$, $\theta$ denotes the coefficient vector,
and $p(\theta | y) \propto h(\theta) = p(y |\theta) p(\theta)$. 
Our VAFC approach will be compared with the DSVI (Doubly Stochastic Variational  Inference) algorithm proposed by \citet{Titsias2014}.  Similar to VAFC, these authors use a 
multivariate normal posterior approximation $q^D (\theta) = N(\mu^D, \Sigma^D)$, where the covariance matrix $\Sigma^D$ is parametrized as $\Sigma^{D} = C^{D} {C^{D}}^T$, with
$C^D$ an unrestricted lower triangular Cholesky factor.  
Both $\mu^D$ and $C^D$ can be updated using a stochastic gradient optimization procedure. The VAFC algorithm differs by parametrizing the covariance matrix through a 
more parsimonious factor structure.  We write $q^F (\theta) = N(\mu^F, \Sigma^F)$ for the VAFC posterior approximation. 

Four examples in Section 4.1 illustrate the performance of DSVI and VAFC when the number of predictors $m$ is moderate and where $m<n$, the kind of situation where there may
be most interest in parameter inference and uncertainty quantification.  We also compare the accuracy of the variational approximations to the exact posterior distribution, computed using MCMC.  The three examples in Section 4.2 consider cases 
in which $m>>n$ and where the 
computational gains from using the factor structure are larger.  
In these saturated models, we employ a horseshoe prior for parameter shrinkage
\citep{Carvalho2010}, so that the variational approximation is to a high-dimensional posterior for both the covariate coefficients and the matching local shrinkage parameters.
In these examples, interest mostly focuses on predictive inference.  Lastly, in Section 4.3 we consider
an example for a mixed effects logistic regression model.  In this case, the variational approximations are to the posterior augmented with a high-dimensional vector of random effect terms. 

In all the examples we set step sizes (learning rates) adaptively using the 
ADADELTA method \citep{Zeiler2012} for both VAFC and DSVI, with different step sizes for each element of $\lambda$.  
Specifically, at iteration $t+1$, the $i$th element $\lambda_i$ of $\lambda$ is updated as 
\begin{align*}
 \lambda_i^{(t+1)}=\lambda_i^{(t)}+\Delta \lambda_i^{(t)}.
\end{align*}
Here, the step size $\Delta \lambda_i^{(t)}$ is $\rho_i^{(t)} g_{\lambda i}^{(t)}$ where $g_{\lambda i}^{(t)}$ denotes the $i$th component of 
$\widehat{\nabla_\lambda {\cal L}(\lambda^{(t)})}$ and $\rho_i^{(t)}$ is
\begin{align*}
  \rho_i^{(t)} & = \frac{\sqrt{E(\Delta_{\lambda i}^2)^{(t-1)}+\epsilon}}{\sqrt{E(g_{\lambda i}^2)^{(t)}+\epsilon}}
\end{align*}
where $\epsilon$ is a small positive constant, with $E(\Delta_{\lambda i}^2)^{(t)}$ and 
$E(g_{\lambda i}^2)^{(t)}$ being decayed running average estimates of ${\Delta \lambda_i^{(t)}}^2$ and ${g_{\lambda i}^{(t)}}^2$, defined by
\begin{align*}
  E(\Delta_{\lambda i}^2)^{(t)} & = \zeta E(\Delta_{\lambda i}^2)^{(t-1)}+(1-\zeta){\Delta \lambda_i^{(t)}}^2 \\
  E(g_{\lambda i}^2)^{(t)} & = \zeta E(g_{\lambda i}^2)^{(t-1)}+(1-\zeta){g_{\lambda i}^{(t)}}^2.
\end{align*}
The variable $\zeta$ is a decay constant.  In the examples we use the default tuning parameter choices $\epsilon=10^{-6}$ and $\zeta=0.95$,
and initialize $E(\Delta_{\lambda i}^2)^{(0)}=E(g_{\lambda i}^2)^{(0)}=0$.  

\subsection{Bayesian logistic regression}

We consider the \texttt{spam}, \texttt{krkp}, \texttt{ionosphere} and \texttt{mushroom} data from the UCI Machine Learning Repository \citep{Lichman2013}. Following \citet{gelman2008}, we change the input matrix into binary variables using the {\tt discretization} function in {\tt R} \citep{Kim2015}.  
After doing this, the \texttt{spam}, \texttt{krkp}, \texttt{ionosphere} and \texttt{mushroom} data respectively contain $n = 4601, 351, 3196$ and $8124$ samples and $m = 104, 111, 37$ and $95$ variables, so that $m<n$ in each case.  In the examples in this section we use a $N(0,10I)$ prior for $\theta$.  

\begin{figure}[!t]
 \centering
\begin{subfigure}{\textwidth}
     \makebox[\textwidth][c]{\includegraphics[width=1.3\textwidth]{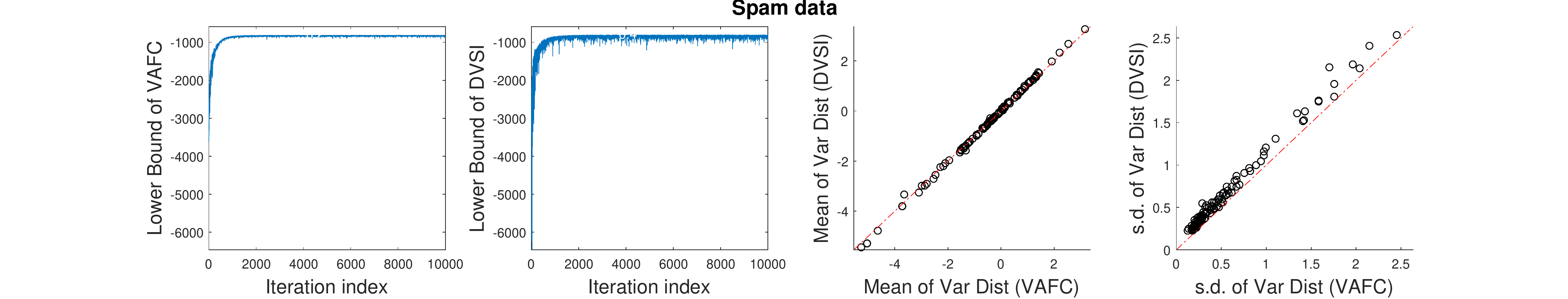}}
     \makebox[\textwidth][c]{\includegraphics[width=1.3\textwidth]{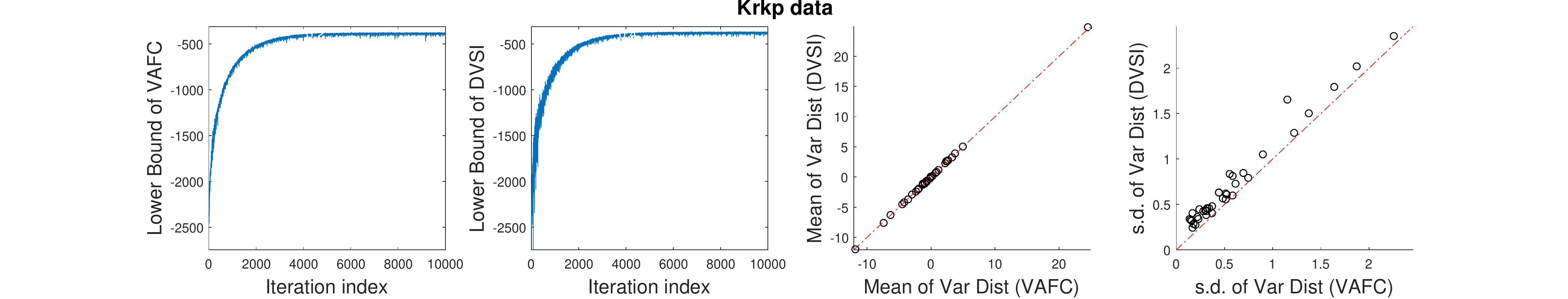}}
     \makebox[\textwidth][c]{\includegraphics[width=1.3\textwidth]{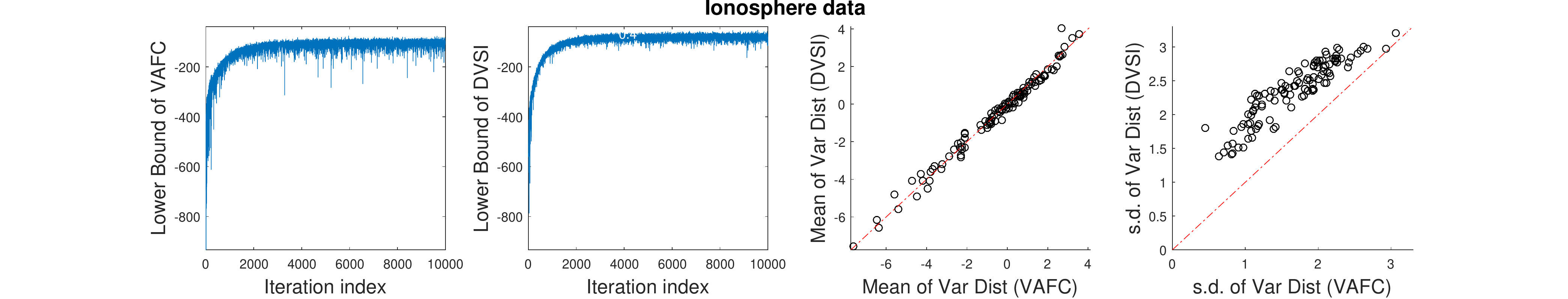}}
     \makebox[\textwidth][c]{\includegraphics[width=1.3\textwidth]{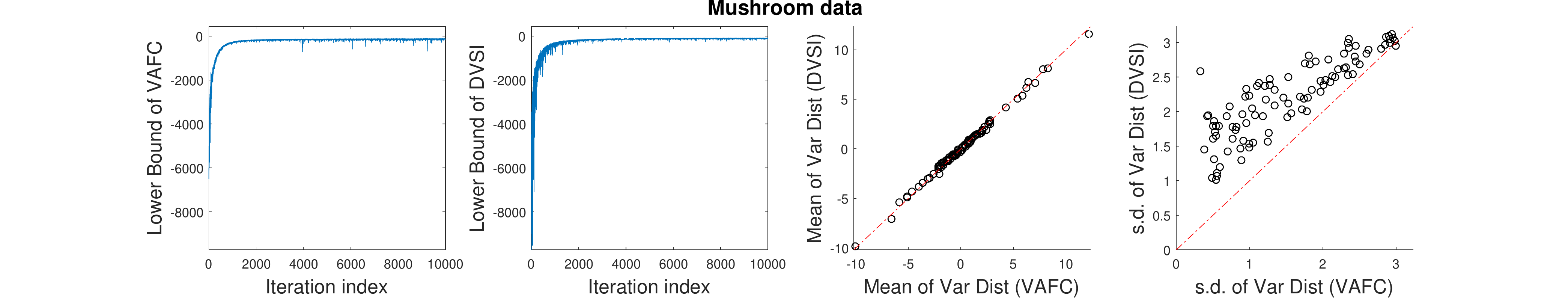}}
\end{subfigure}
  \caption{\label{fig:UCI_log} Monte Carlo estimates of the lower bound, means and standard deviations of the variational distribution of the regression coefficients for both the VAFC with $p=3$ and DVSI approaches. Each row corresponds to a different dataset.  Points near the red lines in the third column indicates that the variational means are similar for both VAFC and DVSI. Points above the red lines in the plots in the last column indicates that the variational standard deviations of the regression coefficients are smaller for VAFC than for DSVI.  }
\end{figure}

\begin{figure}[!t]
  \centering
\begin{subfigure}{\textwidth}
     \makebox[\textwidth][c]{\includegraphics[width=1.3\textwidth]{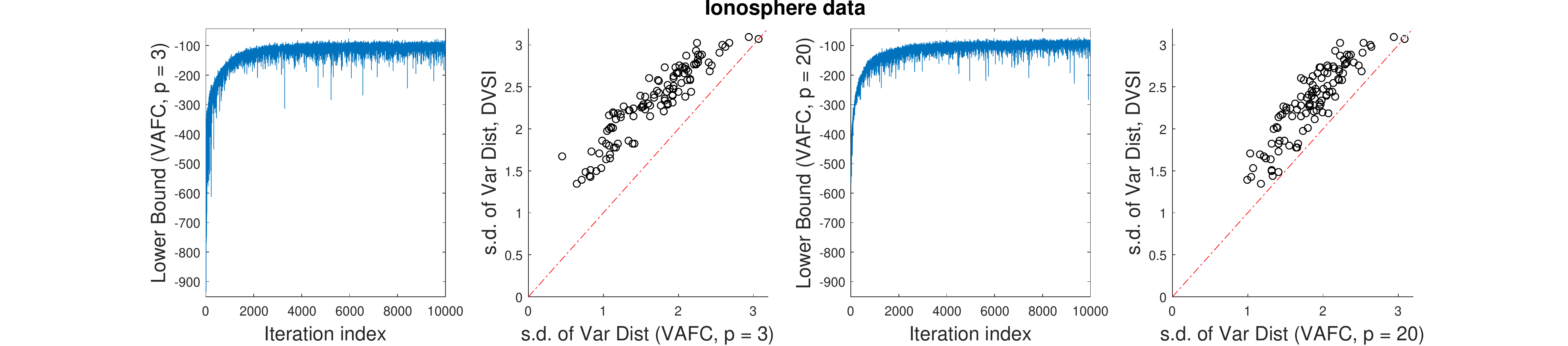}}
     \makebox[\textwidth][c]{\includegraphics[width=1.3\textwidth]{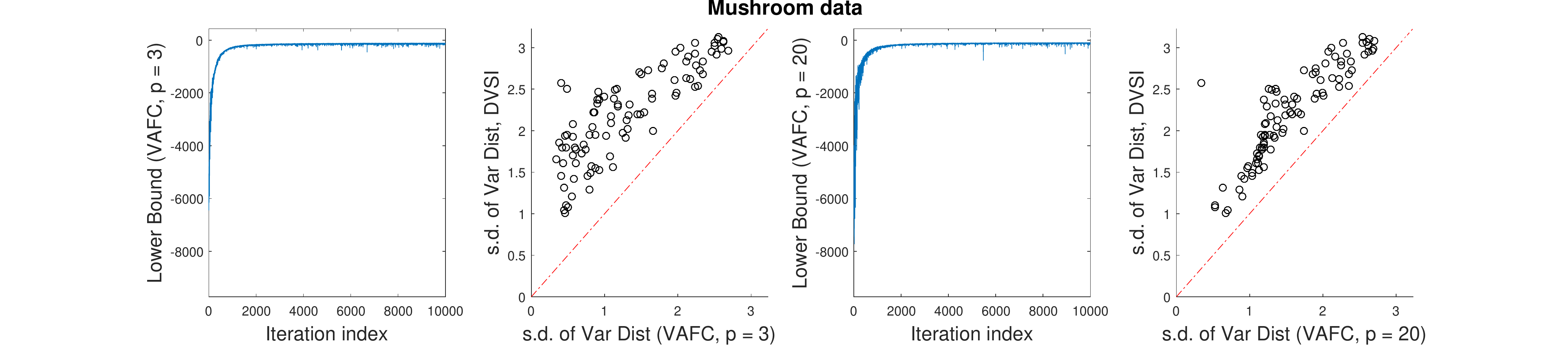}}
\end{subfigure}
  \caption{\label{fig:UCI_log_p20} Monte Carlo estimates of the lower bound, means and standard deviations of the variational distribution of the regression coefficients for both the VAFC with $p=20$ and DVSI approaches. Results are given for the \texttt{ionosphere} and \texttt{mushroom} datsets.  Points near the red lines in the third column indicates that the variational means are similar for both VAFC and DVSI. Points above the red lines in the plots in the last column indicates that the variational standard deviations of the regression coefficients are smaller for VAFC than for DSVI.  }
\end{figure}

\begin{figure}[!h]
  \centering
\begin{subfigure}{\textwidth}
     \makebox[\textwidth][c]{\includegraphics[width=1\textwidth]{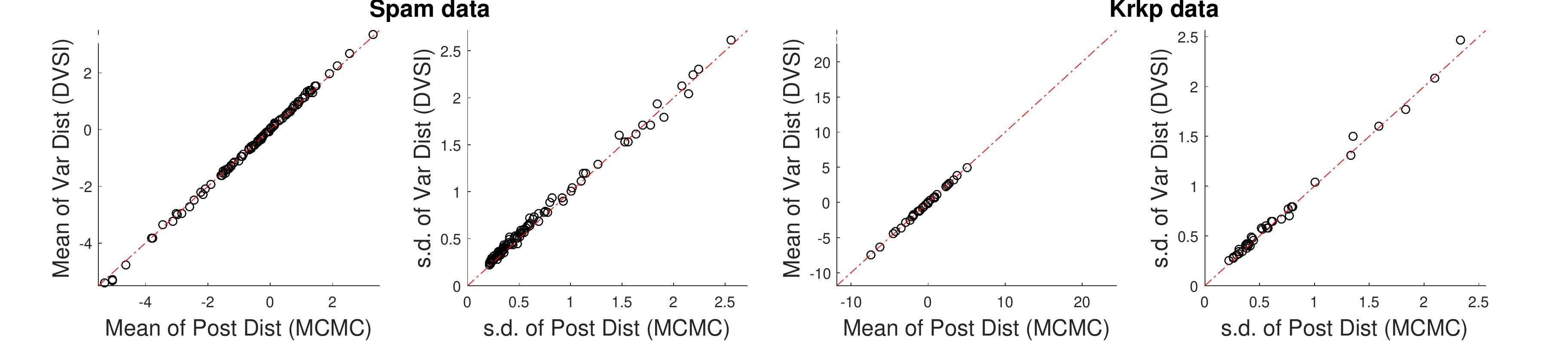}}
     \makebox[\textwidth][c]{\includegraphics[width=1\textwidth]{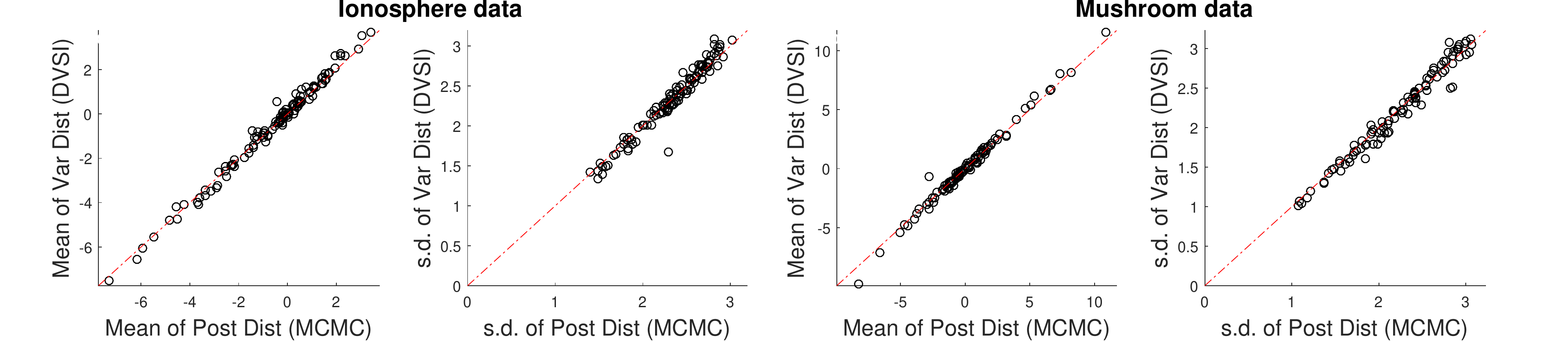}}
\end{subfigure}
  \caption{\label{fig:UCI_MCMC} Scatter plots of the points $(\mu^{MCMC}_{i}, \mu^{D}_{i})$ and $(\sqrt{\Sigma^{MCMC}_{i,i}}, \sqrt{\Sigma^{D}_{i,i}})$ for $i = 1,..,m$ where $\mu^{MCMC}_{i}$ and $\Sigma^{MCMC}_{i,i}$ uses the MCMC approach \texttt{rstanarm}. }
\end{figure}

The first and second columns of Figure \ref{fig:UCI_log} show  respectively Monte Carlo estimates of the lower bounds for DSVI and VAFC with $p=3$ factors over 10,000 iterations.  Convergence is slightly faster for the VAFC method in these examples, and each iteration of the optimization also requires less computation, advantages that are more pronounced in the high-dimensional case considered in Section 4.2.  
To examine the quality of marginal inferences, in the third column of Figure \ref{fig:UCI_log} we plot $(\mu^{F}_{i}, \mu^{D}_{i})$ for $i = 1,...,m$ (i.e. the 
variational means for the two methods) and we see that 
the variational means are close to each other.  The rightmost column of Figure  \ref{fig:UCI_log} shows a similar graphical comparison of the estimated posterior standard deviations of the coefficients for VAFC and DSVI, plotting $(\sqrt{\Sigma^{F}_{i,i}}, \sqrt{\Sigma^{D}_{i,i}})$ for $i = 1,...,m$.  
A variational approximation using an insufficiently flexible approximating family often leads to underestimation of posterior variances (see, for example, \citet{Wang2005}).  
This is indicated here for the VAFC method, with many points appearing above the diagonal lines in the plots.  However, this underestimation of the posterior standard deviations is relatively minor, except for the \texttt{ionosphere} and \texttt{mushroom} datasets.  Figure \ref{fig:UCI_log_p20} shows what happens when the number of factors in the VAFC method is increased to $p=20$ for these datasets and, as expected, this reduces
the underestimation of the standard deviations in the variational posterior.
Although we compare our VAFC method to DSVI in these plots, the DSVI based inferences are very similar to those for the exact posterior computed using MCMC.  
This is illustrated in Figure \ref{fig:UCI_MCMC} where variational posterior means and standard deviations for DSVI are plotted against posterior means and standard deviations computed using MCMC.  For the MCMC computations we used the package {\tt rstanarm} \citep{Stan2016}.  
In this example we have considered results of the VAFC method using $p=3$ and $p=20$ factors.  A reasonable question is how to choose the number of factors in the approximation.  One approach is to calculate the approximation for a sequence of increasing values of $p$, and to stop when posterior inferences of interest no longer change.  
We consider an approach of this kind further in the example of Section 4.3.
\begin{table}[!h]
\begin{center}
\begin{tabular}{|c|cc|cc|} \hline
& \multicolumn{2}{|c|}{VAFC} & \multicolumn{2}{|c|}{DVSI}  \\ \hline
&  Training error &Test Error & Training error & Test Error  \\ \hline
Spam data & 0.046 & 0.058 & 0.046 & 0.057  \\ 
KRKP data & 0.027 & 0.029 & 0.027 & 0.031    \\
Ionosphere data & 0.004 & 0.082 & 0.004 & 0.077   \\
Mushroom data & 0 & 0 & 0 & 0   \\  \hline
\end{tabular}
\end{center}
\caption{\label{table:UCI_log} Average training and test error rates for the four datasets with $m<n$ estimated via five-fold cross validation.  }
\end{table}

Table 1 reports a five-fold cross-validatory assessment of the predictive performance for the four datasets.  For the fitted logistic regressions based on $\mu^D$ and 
$\mu^F$, the average training and test set error rates are very similar for the two approaches. 
This is not surprising given that the variational posterior means tend to be very close for the two methods.  

\subsection{High-dimensional logistic regression examples}

We consider the \texttt{Colon}, \texttt{Leukemia} and \texttt{Breast} cancer datasets available at 
\url{http://www.csie.ntu.edu.tw/~cjlin/libsvmtools/datasets/binary.html}. 
The \texttt{Colon} dataset has $m = 2000$ covariates with sample sizes of $42$ and $20$ in the training and test sets respectively; the \texttt{Leukemia} dataset has $m = 7120$ covariates with sample sizes of $38$ and $34$ in the training and test set; the \texttt{Breast} dataset has similar dimension and sample size as the \texttt{Leukemia} data in the training set, but with only a sample size of $4$ in the test set.  The datasets have $m>>n$ and the posterior distribution is high-dimensional in each case.

\begin{figure}[!t]
  \centering
\begin{subfigure}{\textwidth}
     \makebox[\textwidth][c]{\includegraphics[width=1\textwidth]{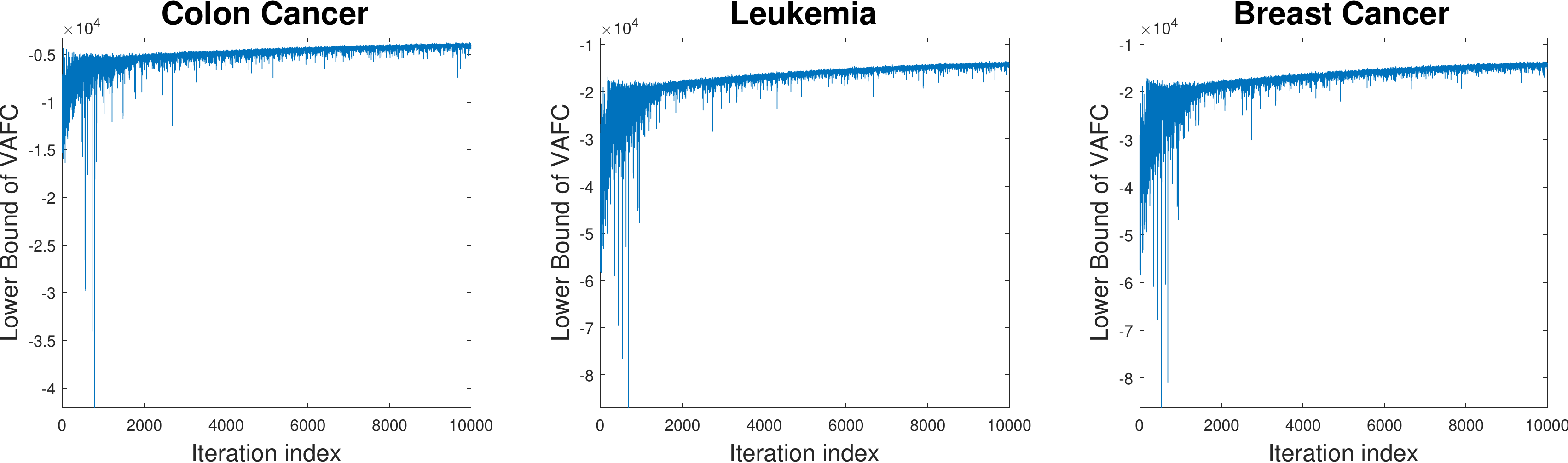}}
\end{subfigure}
  \caption{\label{fig:HighDim_LB} Lower bounds for \texttt{Colon}, \texttt{Leukemia} and \texttt{Breast} cancer data using both the VAFC and DVSI approach.  }
\end{figure}

Here, because of the very high dimensionality of the covariate vectors, we consider a sparse signal shrinkage prior distribution on the coefficients, namely the horseshoe prior 
\citep{Carvalho2010}.  Continuing to write $\theta$ for the regression coefficients as in the last subsection, we now consider the hierarchical prior
$$\theta_j|g,\delta \sim N(0,\delta_j^2 g^2)\;\;\;\delta_j\sim C^+(0,1),$$
for $j=1,\dots,m$, where $C^+(0,1)$ denotes the half-Cauchy distribution.  The parameters $\delta_i$ provide local shrinkage for each coefficient, 
whereas $g$ is a global shrinkage parameter.  For $\theta_0$, we use a $N(0,10)$ prior, and for $g$ we use a half-Cauchy prior, $g\sim C^+(0,1)$.  
We let $v=(v_1,\dots,v_{m+1})^T=(\log \delta_1,\dots,\log \delta_m,\log g)^T$ and denote the full vector of parameters
as $\eta=(\theta^T,v^T)^T$.  We consider a normal variational approximation for $\eta$, using the DSVI and VAFC methods.  
Mean field variational methods are considered for some applications of the horseshoe and other sparse signal shrinkage priors in \citet{Neville2014}.  
Their algorithms do not extend easily to logistic regression, however.

We ran the VAFC algorithm on all three datasets with $p = 4$. Figure \ref{fig:HighDim_LB} shows a Monte Carlo estimate of the lower bound versus iteration number for 10,000 iterations. We found that in this example the DSVI algorithm often diverges even with carefully chosen starting values under our prior settings.   In terms of computation time, using an iMac computer with i5 3.2 Ghz Intel Quad Core, we found that running 100 iterations of VAFC implemented in \texttt{MATLAB} required approximately 32 and 388 seconds for the \texttt{colon} and \texttt{breast} cancer datasets respectively. On the other hand, DVSI required 46 seconds and more than two hours respectively for the same number of iterations and the same datasets. The very slow implementation of DSVI for the \texttt{breast} dataset is related to the memory requirements of the DSVI approach, which is another relevant aspect of the comparison of the algorithms.  
Note that the timings presented are for the same fixed number of iterations, and the reduced number of variational parameters in the VAFC approach often means than the number of iterations required for convergence is much reduced, so the the reduction in computation time is substantial for the VAFC method.  

In these high-dimensional examples \citet{Titsias2014} considered a version of their procedure using a diagonal covariance matrix and a feature selection approach
based on automatic relevance determination (DSVI-ARD).  We compare predictive performance of the DSVI-ARD approach with the VAFC method with $p=4$ factors and the horseshoe prior
in Table \ref{table:cancer_error}.  The DSVI-ARD results are those reported in \citet{Titsias2014}.  Similar predictive performance is achieved by the two methods.
\begin{table}[!h]

\begin{center}
\begin{tabular}{|c|cc|cc|} \hline
& \multicolumn{2}{|c|}{VAFC} & \multicolumn{2}{|c|}{DVSI-ARD}  \\ \hline
&  Training error &Test Error & Training error & Test Error  \\ \hline
\texttt{Colon} & 0/42 & 0/20 & 0/42 & 1/20  \\ 
\texttt{Leukemia} & 0/38 & 6/34 & 0/38 & 3/34    \\
Ionosphere data & 0/38 & 1/4 & 0/38 & 2/4 \\ \hline
\end{tabular}
\end{center}

\caption{\label{table:cancer_error} Train and test error rates for the three cancer datasets for the VAFC and DVSI-ARD methods. Errors rates are reported as the ratio of misclassified data points over the number of data points. }
\end{table}

\subsection{Mixed logistic regression}

In this example, we consider a random intercept model for the polypharmacy data set described in \citet{Hosmer2013}. This longitudinal dataset is available at \url{http://www.umass.edu/statdata/statdata/stat-logistic.html}, and contains data on 500 subjects, who were followed over seven years. 
Following \citet{Tan2016}, we consider a logistic mixed effects model of the form
\begin{align}\label{poly_model}
\text{logit } p(y_{ij} = 1 | \theta) &= \beta_0 + \beta_{\texttt{gender}} \text{Gender}_i +  \beta_{\texttt{race}} \text{Race}_i + \beta_{\texttt{age}} \text{Age}_{ij} \nonumber \\ 
& + \beta_{M1} \text{MHV1}_{ij}  + \beta_{M2} \text{MHV2}_{ij}  + \beta_{M3} \text{MHV3}_{ij} \\
&+ \beta_{IM} \text{INPTMHV}_{ij} + u_i \nonumber 
\end{align}
for $i = 1,2,...,500$ and $j = 1,2,...,7$.  The response variable $y_{ij}$ is $1$ if subject $i$ in year $j$ is taking drugs from three or more different classes, and $-1$ otherwise. The covariate $\text{Gender}_i = 1$ if subject $i$ is male and 0 if female; $\text{Race}_i = 0$ if the race of subject $i$ is white and $1$ otherwise; and letting $\text{MHV}_{ij}$ be the number of outpatient mental health visits for subject $i$ and year $j$, we set $\text{MHV1}_{ij} = 1$ if $1 \leq \text{MHV}_{ij} \leq 5$ and 0 otherwise,  $\text{MHV2}_{ij} = 1$ if $6 \leq \text{MHV}_{ij} \leq 14$ and 0 otherwise, and $\text{MHV3}_{ij} = 1$ if $\text{MHV}_{ij} \geq 15$ and 0 otherwise. The covariate $\text{INPTMHV}_{ij}$ is $0$ if there were no inpatient mental health visits for subject $i$ in year $j$ and 1 otherwise. Finally $u_i\sim N(0,\exp(2\zeta))$ is a subject level random intercept. 
Write $\beta = (\beta_0, \beta_{\texttt{gender}}, \beta_{\texttt{race}}, \beta_{\texttt{age}} , \beta_{M1}, \beta_{M2} , \beta_{M3}, \beta_{IM})^T$, $u=(u_1,\dots,u_{500})^T$ 
and the parameters augmented with the random intercepts as $\theta=(\beta^T,u^T,\zeta)^T$.  
The prior distribution takes the form 
\[
p(\theta) = p(\beta) p(\zeta) \prod_{i=1}^n p(u_i | \zeta)
\]
where $p(\beta)$ is $N(0,100 I_8)$, $p(\zeta)$ is $N(0,100)$ and $p(u_i|\zeta)$ is $N(0,\exp(2\zeta))$.

\begin{figure}[!ht]
  \centering
\begin{subfigure}{\textwidth}
     \makebox[\textwidth][c]{\includegraphics[width=0.5\textwidth]{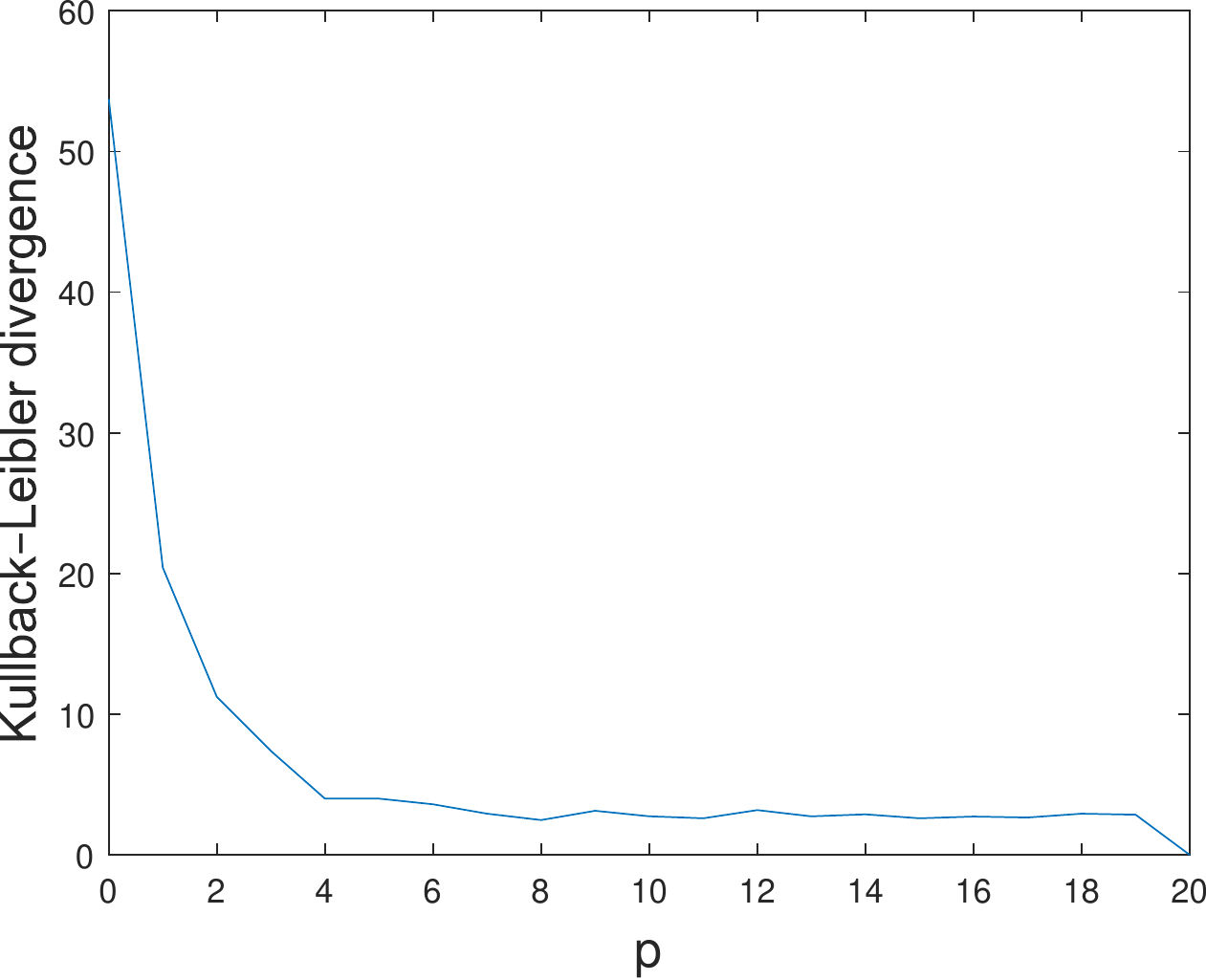}}
\end{subfigure}
  \caption{\label{fig:poly_KL} Kullback-Leibler divergence between the final distribution of VAFC for multiple values of $p$ and VAFC using $p = 20$.}
\end{figure}

\begin{figure}[!ht]
  \centering
\begin{subfigure}{\textwidth}
     \makebox[\textwidth][c]{\includegraphics[width=1.3\textwidth]{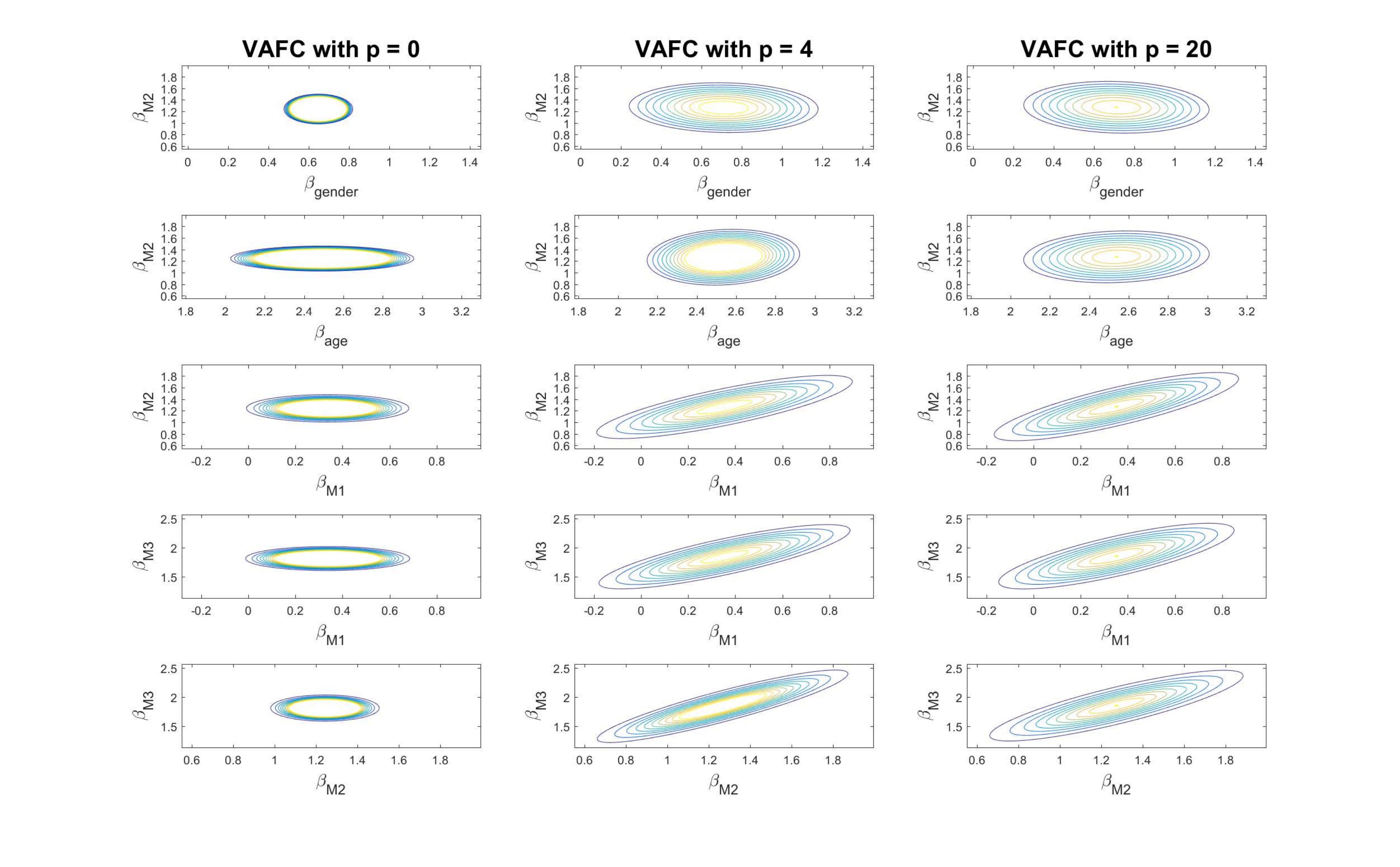}}
\end{subfigure}
  \caption{\label{fig:poly_contour} Bivariate contour plots of the posterior density of coefficients with the five highest correlations for the VAFC.}
\end{figure}

We ran the VAFC algorithm for 10,000 iterations using $p = 0,1,...20$ factors.  Figure \ref{fig:poly_KL} shows the KL divergence between the variational distribution with $p$ factors and that with $20$ factors as $p$ varies (note that the KL divergence between two multivariate Gaussian distributions is computable in closed form).  This shows that the variational approximation to the posterior augmented with the random effects is similar for $p\geq 4$. 
To illustrate this further, Figure \ref{fig:poly_contour} shows contour plots of some selected bivariate variational posterior marginals.  
The results when $p=0$ (i.e. a diagonal approximation) are very different, and 
even a crude allowance for posterior correlation with a small number of factors can grealy improve estimation of 
the posterior marginal distributions.
\begin{figure}[!ht]
  \centering
\begin{subfigure}{\textwidth}
     \makebox[\textwidth][c]{\includegraphics[width=1\textwidth]{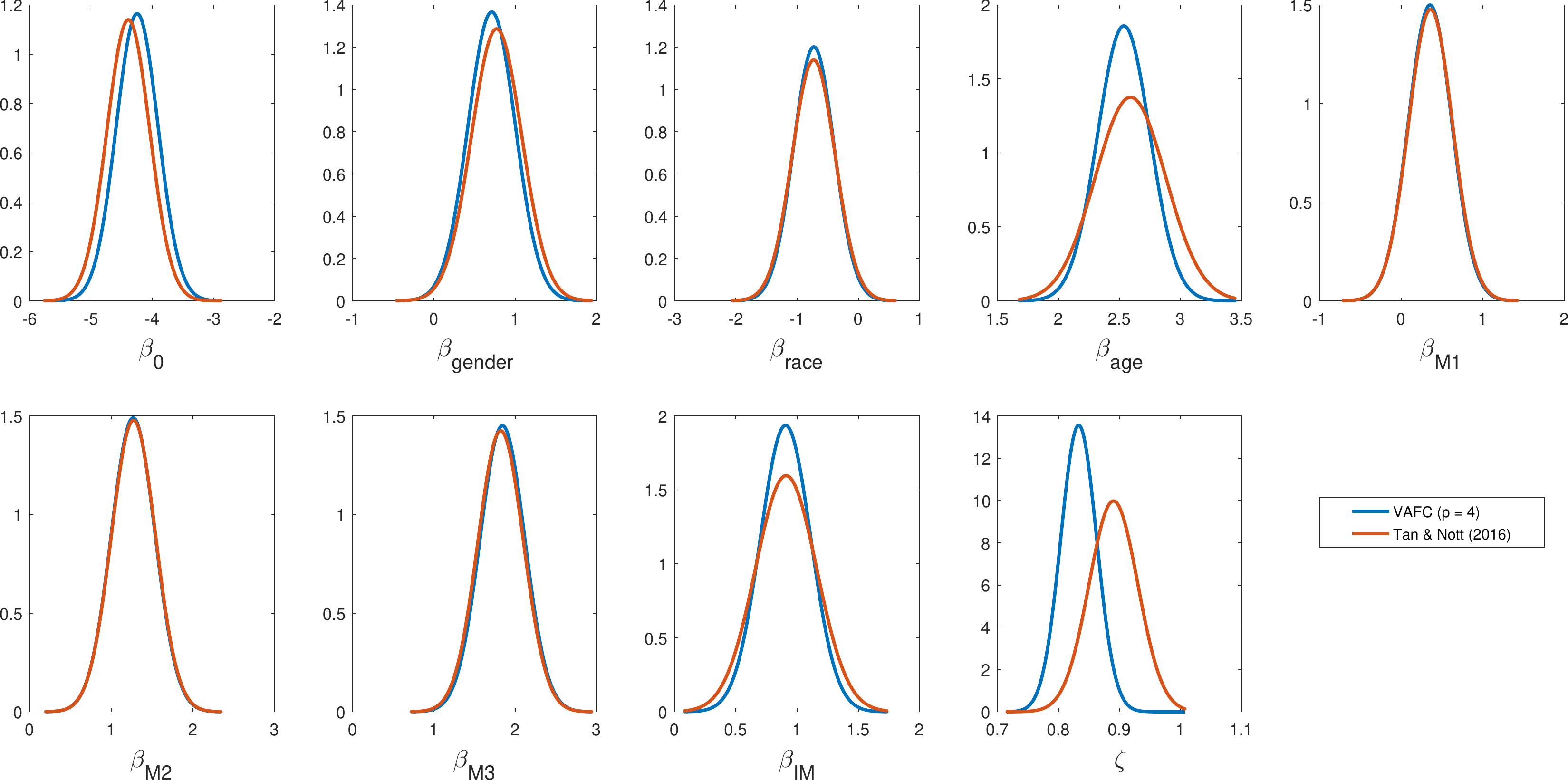}}
\end{subfigure}
  \caption{\label{fig:poly_tan} Marginal posterior distributions of components of $\beta$ and $\zeta$ using the method of \citet{Tan2016} and VAFC with $p=4$.}
\end{figure}
Finally, we also compare the variational marginal density of the regression coefficients with the method in \citet{Tan2016}.  
The method of \citet{Tan2016} gives similar answers to MCMC in this example, as shown in Figure 5 of their manuscript, so the \citet{Tan2016} can be considered
both a gold standard for a normal approximation as well as a good gold standard more globally.  
Figure \ref{fig:poly_tan} shows that, except for some mild underestimation of the random intercept variance parameter $\zeta$, the VAFC algorithm with $p = 4$ provides good approximations of the marginal posterior distributions of the components of $\beta$. 
Figure \ref{fig:random_int_compare} shows plots of the variational posterior means and standard deviations of the subject level random intercepts for VAFC with $p=4$ against those for the method of  
\citet{Tan2016}.  The posterior distributions of random intercepts are close for the two methods.  
\begin{figure}[!ht]
  \centering
\begin{subfigure}{\textwidth}
     \makebox[\textwidth][c]{\includegraphics[width=0.8\textwidth,height=0.4\textwidth]{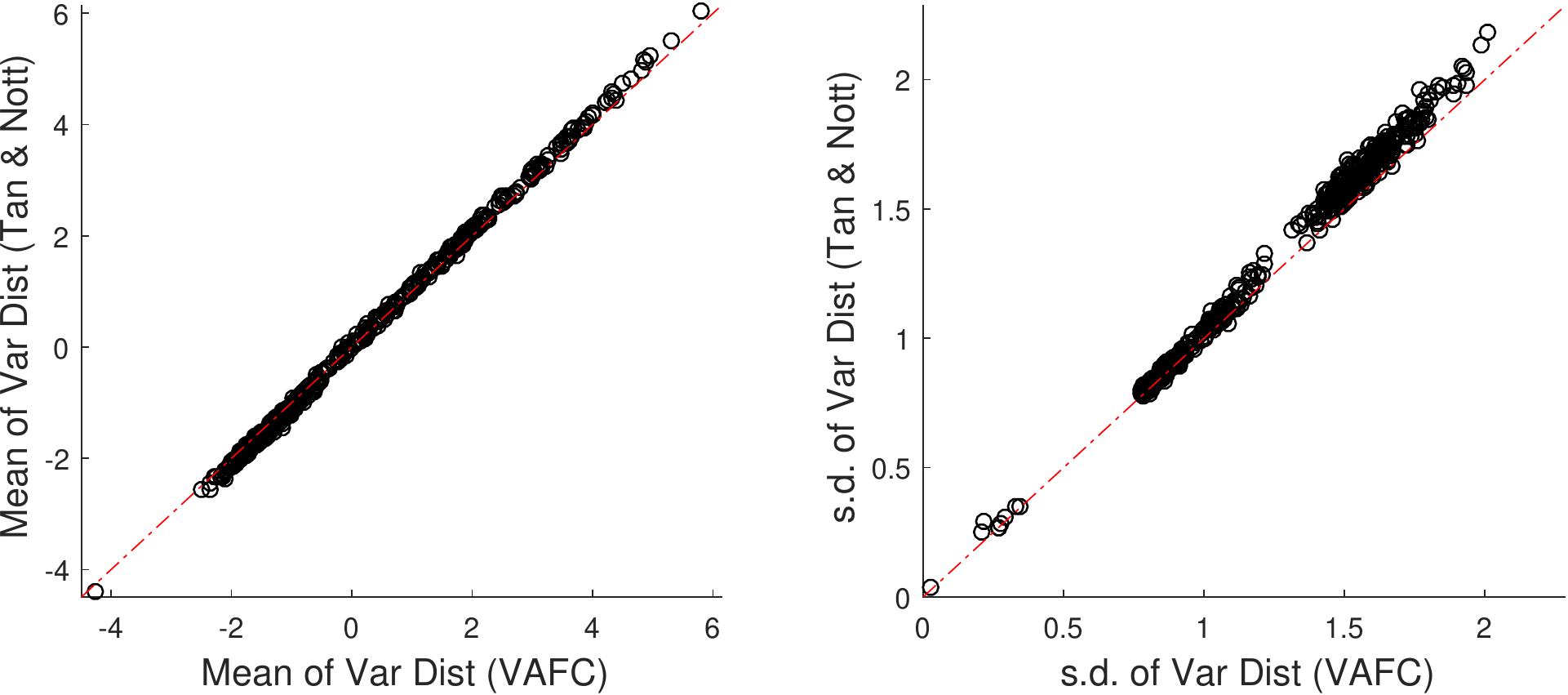}}
\end{subfigure}
\caption{\label{fig:random_int_compare} Plot of variational posterior means (left) and standard deviations (right) of the subject level random intercepts for VAFC with $p=4$ against those for the method of Tan and Nott}
\end{figure}

\section{Discussion}

To construct practical variational approximation methods in high dimensions it is important to employ parsimonious but flexible parametrizations of variational families.  Gaussian approximations are important, both because they are useful in themselves, but also as a building block for more sophisticated approaches such as variational mixture
approximations \citep{Jaakkola1998,gershman+hb12,Salimans2013,Guo2016,Miller2016} or approximations based on Gaussian copulas \citep{Han2016}.  
Here we have considered factor covariance structures for Gaussian variational approximation in situations where
there is no natural conditional independence structure that can be exploited in the model for reducing the number of free covariance parameters.  
The approximations can be efficiently formed using the reparametrization trick for gradient estimation and exploiting the Woodbury formula to compute the gradient estimates.  
In applications to logistic regression and generalized linear mixed models the methods perform very well.   

One difficulty in application of the presented method relates to the problem of choosing a suitable number of factors.  As mentioned in the examples, a useful and obvious heuristic is to apply the method for an increasing sequence of
values of $p$ and to stop when inferences of interest no longer change.  
In applications where a higher level of accuracy is needed it will be important to go beyond Gaussian approximations of the type considered here, such as using mixture or copula approximations and the recently developed variational boosting approaches of \citet{Guo2016} and \citet{Miller2016} may be particularly useful in this respect.  It is also possible in the Gaussian case to combine factor structure with knowledge of relevant conditional independence relationships in the model.  There is room for much ingenuity in exploiting the
structure of the model itself for suggesting parsimonious and expressive parametrizations of variational families for particular applications.  

\section*{Acknowledgements}

David Nott and Victor Ong were supported by a Singapore Ministry of Education Academic Research Fund Tier 2 grant (R-155-000-143-112).  We thank Linda Tan for helpful comments on an earlier draft of the manuscript.  

\newpage

\section*{Appendix - derivation of gradient expressions}

In this subsection we give a derivation of the gradient expressions (\ref{gradmu})-(\ref{gradd}).
We consider gradients for each term in (\ref{lowerbound}) separately.  
We will make use of the following identity.  
If $A$, $B$ and $C$ are conformably dimensioned matrices, then
$\mbox{vec}(ABC)=(C^T\otimes A)\mbox{vec}(B)$, where $\otimes$ denotes the Kronecker product.  
Looking at the first term on the right in (\ref{lowerbound})
\begin{align*}
  \nabla_\mu E_f(\log h(\mu+Bz+d\circ \epsilon)) & = E_f(\nabla_\theta \log h(\mu+Bz+d\circ\epsilon)), \\
  \nabla_{\mbox{vec$(B)$}} E_f(\log h(\mu+Bz+d\circ\epsilon)) & = E_f(\nabla_{\mbox{vec$(B)$}} \log h(\mu+(z^T\otimes I)\mbox{vec}(B)+d\circ \epsilon)) \\
 & = E_f((z^T\otimes I)^T \nabla_\theta \log h(\mu+Bz+d\circ \epsilon)) \\
 & = E_f((z\otimes I) \nabla_\theta \log h(\mu+Bz+d\circ\epsilon)) \\
 & = \mbox{vec}(E_f(\nabla_\theta \log h(\mu+Bz+d\circ \epsilon) z^T))
\end{align*}
or $\nabla_B E_f(\log h(\mu+Bz+d\circ\epsilon))=E_f(\nabla_\theta \log h(\mu+Bz+d\circ \epsilon)z^T)$.
Finally, writing $d\circ \epsilon=D\epsilon$ and noting the symmetry of the way that $Bz$ and $D\epsilon$ appear in the above expression we can
write
\begin{align*}
 \nabla_D E_f(\log h(\mu+Bz+d\circ\epsilon))= & E_f(\nabla_\theta \log h(\mu+Bz+d\circ\epsilon)\epsilon^T)
\end{align*}
which gives $\nabla_d E_f(\log h(\mu+Bz+d\circ\epsilon))=  E_f(\mbox{diag}(\nabla_\theta \log h(\mu+Bz+d\circ\epsilon) \epsilon^T))$.

The second term on the right hand side of (\ref{lowerbound}) is constant in the variational parameters and hence can be neglected.  
Next, consider the third term.   Here we use the following results from matrix calculus (see, for example, \citet{Magnus1999}).  
For a square invertible matrix $A$, $\nabla_{A} \log |A| = A^{-1}$.  Also, for $A$ a $m\times p$ matrix, 
write $\frac{\mbox{$d$ vec$(AA^T)$}}{\mbox{$d$ vec$(A)$}}$ for the $m^2\times mp$ matrix where the $(i,j)$th entry is
the derivative of the $i$th entry of $\mbox{vec}(AA^T)$ with respect to the $j$th entry of $\mbox{vec}(A)$.  Then 
$$\frac{d \mbox{vec}(AA^T)}{d \mbox{vec}(A)}=(I+K_{mm})(A\otimes I)$$
where $K_{pm}$ is the commutation matrix \citep{Magnus1999} of dimensions $pm\times pm$ which satisfies $K_{pm}\mbox{vec}(A)=\mbox{vec}(A^T)$.  
A useful property of the commutation matrix we will need later is the following.  If $A$ is a $p\times m$ matrix, and $C$ is an $r\times s$ matrix, then 
$K_{pr}(A\otimes C)=(C\otimes A) K_{ms}$.  
We have 
\begin{align*}
  \nabla_\mu E_f\left( \frac{1}{2}\log |BB^T+D^2|\right) & = 0,
\end{align*}
\begin{align*}
  \nabla_{\mbox{vec$(B)$}} E_f\left(\frac{1}{2}\log |BB^T+D^2|\right)= & \frac{1}{2}\left\{(I+K_{mm})(B\otimes I)\right\}^T \mbox{vec}((BB^T+D^2)^{-1}) \\
 = & \frac{1}{2}\left\{(B^T\otimes I)\mbox{vec}((BB^T+D^2)^{-1})+\right. \\
 & \hspace{0.5in}\left. (B^T\otimes I)K_{mm}\mbox{vec}((BB^T+D^2)^{-1})\right\} \\
 = & \mbox{vec}((BB^T+D^2)^{-1} B)
\end{align*}
and hence $\nabla_B E_f\left(\frac{1}{2}\log |BB^T+D^2|\right)=(BB^T+D^2)^{-1} B$.  
Again noting the symmetry of the way that $BB^T$ and appear we have
$\nabla_d E_f\left(\frac{1}{2}\log |BB^T+D^2|\right) = \mbox{diag}((BB^T+D^2)^{-1} D)$.  

Finally, consider the last term on the right of $(\ref{lowerbound})$.  We need the following product rule from matrix differential calculus (again we refer
the reader to \citet{Magnus1999}).  If $g(A)$ and $k(A)$ are matrix-valued functions, conformably dimensioned, of the matrix $A$, then
\begin{align*}
  \nabla_A \mbox{tr}(f(A)^T k(A)) & =\left.\left\{\nabla_A \mbox{tr}(f(A)^T  k(C))+\nabla_A \mbox{tr}(k(A)^T f(C))\right\}\right|_{C=A}.
\end{align*}
Using this result
\begin{align}
 \nabla_B E_f\left(\frac{1}{2}\mbox{tr}((Bz+d\circ\epsilon)^T (BB^T+D^2)^{-1}(Bz+d\circ\epsilon))\right) & = \frac{1}{2}E_f\left(T_1+T_2\right) \label{gradBcalc}
\end{align}
where
\begin{align*}
 T_1=& \left.\left\{\nabla_B \mbox{tr}((Bz+d\circ\epsilon)(Bz+d\circ\epsilon)^T CC^T+D^2)^{-1}\right\}\right|_{C=B}, \nonumber \\
 T_2 =& \left.\left\{\nabla_B \mbox{tr}((BB^T+D^2)^{-1}(Cz+d\circ \epsilon)(Cz+d\circ\epsilon)^T)\right\}\right|_{C=B}. \nonumber 
\end{align*}
Evaluating $T_1$ gives
\begin{align*}
 T_1 & =  \left.\mbox{vec}^{-1}(\nabla_{\mbox{vec$(B)$}} ((z^T\otimes I)\mbox{vec}(B)+d\circ\epsilon)^T (CC^T+D^2)^{-1} ((z^T\otimes I)\mbox{vec}(B)+d\circ\epsilon))\right|_{C=B} \\
 & = \left.\mbox{vec}^{-1}( 2(z\otimes I)(CC^T+D^2)^{-1}((z^T\otimes I)\mbox{vec}(B)+d\circ\epsilon))\right|_{C=B} \\
 & = 2(BB^T+D^2)^{-1}(Bz+d\circ\epsilon)z^T.
\end{align*}

To evaluate $T_2$, we need one further result.  Write 
\begin{align*}
 & \frac{\mbox{$d$ vec$(A^{-1})$}}{\mbox{$d$ vec$(A)$}}
\end{align*} 
for the matrix with $(i,j)$th entry given by the derivative of 
the $i$th entry of $\mbox{vec$(A^{-1})$}$ with respect to the $j$th entry of $\mbox{vec}(A)$.  Then
\begin{align*}
\frac{\mbox{$d$ vec}(A^{-1})}{\mbox{$d$ vec}(A)} & = -(A^{-T}\otimes A^{-1})
\end{align*}
We have
\begin{equation*}
\begin{split}
 T_2 =&  \left. \mbox{vec}^{-1}(\nabla_{\mbox{vec$(B)$}} (Cz+d\circ \epsilon)^T(BB^T+D^2)^{-1}(Cz+d\circ\epsilon))  \right|_{C=B} \\
 = & \left.\mbox{vec}^{-1}(\nabla_{\mbox{vec$(B)$}}(Cz+d\circ\epsilon)^T\left\{(Cz+d\circ\epsilon)^T\otimes I\right\}\mbox{vec}((BB^T+D^2)^{-1}))\right|_{C=B} \\
 = & -\left\{\left\{\frac{\mbox{$d$ vec$(BB^T+D^2)$}}{\mbox{$d$ vec$(B)$}}\right\}^T  \left\{(BB^T+D^2)^{-1}\otimes (BB^T+D^2)^{-1}\right\} \right. \\
  & \hspace{0.5in} \left.\left.\left.\left\{(Cz+d\circ\epsilon)\otimes I\right\} (Cz+d\circ\epsilon)\right\}\vphantom{\left\{\frac{\mbox{$d$ vec$(BB^T+D^2)$}}{\mbox{$d$ vec$(B)$}}\right\}^T}\right\}\right|_{C=B} \\
  = & -\left.\mbox{vec}^{-1}((B^T\otimes I) (I+K_{qq})  (BB^T+D^2)^{-1}\otimes(BB^T+D^2)^{-1} \right. \\
   & \hspace{0.5in}\left. (Cz+d\circ\epsilon)\otimes I (Cz+d\circ\epsilon)\right|_{C=B} \\
  = & -\left.\mbox{vec}^{-1}( (B^T\otimes I) (BB^T+D^2)^{-1}\otimes (BB^T+D^2)^{-1} (Cz+d\circ\epsilon)\otimes I (Cz+d\circ\epsilon) )\right|_{C=B} \\
    & -\left. \mbox{vec}^{-1}(K_{qq} (I\otimes B^T)  (BB^T+D^2)^{-1}\otimes (BB^T+D^2)^{-1} (Cz+d\circ\epsilon)\otimes I (Cz+d\circ\epsilon))\right|_{C=B} \\
   = & -\left.\mbox{vec}^{-1}(B^T(BB^T+D^2)^{-1}(Cz+d\circ\epsilon)\otimes (BB^T+D^2)^{-1})\right|_{C=B} \\
   & -\left.\mbox{vec}^{-1}(K_{qq} (BB^T+D^2)^{-1}(Cz+d\circ\epsilon)\otimes B^T(BB^T+D^2)^{-1} (Cz+d\circ\epsilon))\right|_{C=B} \\
   = & -2\left.\mbox{vec}^{-1}(\mbox{vec}(B^T(BB^T+D^2)^{-1}(Cz+d\circ\epsilon)(Cz+d\circ\epsilon)^T(BB^T+D^2)^{-1}))\right|_{C=B} \\
   = & -2(BB^T+D^2)^{-1}(Bz+d\circ\epsilon)(Bz+d\circ\epsilon)^T(BB^T+D^2)^{-1}B 
\end{split}
\end{equation*}
Hence the required expression at (\ref{gradBcalc}) is
\begin{align*}
 \frac{1}{2}E_f(T_1+T_2) = &  E_f((BB^T+D^2)^{-1}(Bz+d\circ\epsilon)z^T \\
  & -(BB^T+D^2)^{-1}(Bz+d\circ\epsilon)(Bz+d\circ\epsilon)^T(BB^T+D^2)^{-1}B).
\end{align*}
Again noting the symmetry in the way that $B$ and $D$ appear there is immediately a similar expression to (\ref{gradBcalc}) for the gradient with respect to $D$, 
and taking the diagonal gives the appropriate gradient with respect to the vector $d$ of diagonal elements.

Collecting all the previous results together for the terms in the lower bound (\ref{lowerbound}) gives the gradient expressions (\ref{gradmu})-(\ref{gradd}).

\bibliographystyle{chicago}
\bibliography{ref}

\end{document}